\documentclass[journal,twoside]{IEEEtran}
\usepackage{amsmath,amsfonts}
\usepackage[caption=false,font=normalsize,labelfont=sf,textfont=sf]{subfig}
\usepackage{stfloats}
\usepackage{url}
\usepackage{verbatim}
\usepackage{graphicx}
\usepackage{cite}
\usepackage[utf8]{inputenc}

\begin{document}

\title{From Surface Lakes to Subsurface Signatures in Antarctica with BIOMASS P-Band TomoSAR}

\author{Francesco Banda
\thanks{October 29, 2026. (Corresponding author: Francesco Banda). Francesco Banda
is with Aresys, Vimodrone (Milan), Italy.}}

\markboth{Geoscience and Remote Sensing, August~2026}%
{Banda: From Surface Lakes to Subsurface Signatures in Antarctica with BIOMASS P-Band TomoSAR}

\IEEEpubid{This work has been submitted to the IEEE for possible publication. Copyright may be transferred without notice, after which this version may no longer be accessible.}

\maketitle

\begin{abstract}
Spaceborne remote sensing of the cryosphere is fundamental to map ice structure and dynamics. Accurate mass balance estimation requires a better understanding of melt/refreeze processes and firn hydrology, whereas spaceborne optical and Radar sensors are mostly limited to near-surface observations. The BIOMASS P-band Synthetic Aperture Radar, launched by European Space Agency in 2025, provides an unprecedented opportunity to gain a more direct access to englacial processes, thanks to the \mbox{70 cm} wavelength paired with polarimetry, interferometry and tomography. This paper investigates subsurface beneath a region of historical supraglacial lake activity with BIOMASS tomography, at the Riiser-Larsen Ice Shelf in East Antarctica. The main result is that subsurface scattering appears to be related to historical supraglacial lake recurrence, with structural and depth variations over time. This suggests that recurrent supraglacial lake activity may leave persistent hydrological signatures within the firn. Overall, this study highlights the potential of gaining valuable insights into ice sheet hydrology with P-band.
\end{abstract}

\begin{IEEEkeywords}
Synthetic Aperture Radar, P-band, Interferometry, Tomography, BIOMASS, Ice Sheet, Antarctica, Meltwater.
\end{IEEEkeywords}

\section{Introduction}

\IEEEPARstart{I}{ce} sheets and glaciers play a fundamental role in regulating climate, global water circulation and sea level. Polar ice sheets altogether store the vast majority of world's freshwater, Antarctica being particularly relevant due to its vast mass and potential impact on sea level rise. Investigating ice structure and dynamics is thus a scientific priority, for which spaceborne remote sensing provides a practical solution with large-scale and periodic monitoring \cite{lenaerts2019observing}.\\
Accurate mass balance estimates require better understanding of firn hydrology, particularly meltwater percolation and refreezing. Firn retains meltwater within its pores, reducing surface runoff. However, increasing melt rates and ice layer formation can reduce its storage capacity. Improved understanding is therefore critical for firn evolution models and reliable projections of mass balance and sea level rise \cite{lenaerts2019observing}.\\
Optical, altimetric and Synthetic Aperture Radar (SAR) spaceborne sensors mostly provide near-surface observations, offering limited information on englacial processes. Consequently, mass balance, firn density and ice thickness are often estimated indirectly \cite{lenaerts2019observing}. While subsurface sensing was previously limited to in-situ surveys \cite{sharma2012estimation, banda2015single, hubbard2016massive, montgomery2020hydrologic}, a change of paradigm is represented by BIOMASS, the first spaceborne P-band SAR, launched by the European Space Agency (ESA) in 2025 \cite{scipal2026biomass}. The wavelength of about 70 cm paired with polarimetry, interferometry (InSAR) and tomography (TomoSAR), enables mapping of the englagial structure in Antarctica, with resolution and performance depending on mission phase \cite{banda2026eusar}.\\
Following initial exploratory work \cite{banda2026eusar}, this manuscript uses BIOMASS TomoSAR to investigate subsurface beneath a region of historical supraglacial lake activity. Supraglacial lakes are important in Antarctic ice shelf hydrology, as they provide temporary storage for meltwater, which can infiltrate into the firn. Subsequent refreezing can alter firn density, thermal properties, and internal structure, while also producing features that may persist beyond the melt season. This can influence the firn's capacity to store meltwater in subsequent years. The studied site is the highly relevant Riiser-Larsen Ice Shelf (RLIS) in East Antarctica, characterized by substantial interannual variability in supraglacial lake activity \cite{dirscherl2021seasonal, arthur2022large} and recent evidence of water within the firn \cite{frank2026mysterious}.\\
The manuscript is structured as follows: Section \ref{sec:methodology} presents the methodology applied to BIOMASS stacks, highlighting the importance of ad-hoc phase calibration; Section \ref{sec:dataset} describes the dataset used in this study; Section \ref{sec:results} discusses experimental results; Sections \ref{sec:discussion} and \ref{sec:conclusions} contain discussion and conclusions.

\IEEEpubidadjcol

\section{Methodology}
\label{sec:methodology}

\subsection{InSAR and TomoSAR Processing}

To generate BIOMASS TomoSAR stacks, data are first coregistered and ionospherically calibrated as in \cite{banda2026eusar, tebaldini2026multi}, where the choice between standard coregistration and multi-squint InSAR processing, for each pair, depends on the severity of ionosphere. The stack is subsequently phase calibrated at TomoSAR level by estimating \textit{surface phases}, expressed as:

\begin{equation}
	\varphi^{n}(x,r)= k_z^n(x,r)z_{C}(x,r)+\alpha^n(x,r)
	\label{eq:steering}
\end{equation}

\noindent where: $(x,r)$ is the azimuth-range cell; $n$ is the acquisition index; $k_z^n$ is the InSAR wavenumber; $\alpha^n$ is residual low-pass disturbance \cite{banda2026eusar}; $z_{C}$ is surface topography.\\
Surface phases are here estimated as follows: Sum of Kronecker Products Decomposition (SKPD) \cite{banda2015single, tebaldini2011multibaseline} is used to separate InSAR surface and subsurface contributions from the polarimetric-interferometric stack coherence matrix, where the four boundary solutions are retained, i.e., most and least coherent surface and subsurface within the interval of physical validity \cite{tebaldini2011multibaseline}; Singular Value Decomposition (SVD) is computed on the four resulting InSAR matrices, the solution closest to rank-one approximation is selected and the calibration phases are given by the phases of its first singular vector; surface phases are finally composed by the calibration phases, additionally steered with TomoSAR beamforming to track the highest scattering mechanism of the four. The application of surface phases to the ionospherically calibrated stack ideally results in removal of residual disturbances and surface steering to zero elevation.\\
Alternative phases are estimated with Phase Linking (PL) on HH polarization for comparison \cite{guarnieri2008exploitation}. For PL $z_{C}$ does not in general correspond to surface \cite{banda2026eusar}, but rather to the effective vertical scattering center. While additional surface steering may be required, here $z_{C}$ retrieved with PL and SKPD is directly compared.\\
Finally, TomoSAR 3D reconstruction is computed \cite{banda2026eusar}. Although having lower vertical resolution, Fourier beamforming is used as the baseline estimator, motivated by the more conservative reconstruction of vertical scattering to compare different polarimetric channels and stacks. Capon super-resolution estimator is used where needed, to partially improve the Fourier coarser response for analysis.

\subsection{Expected Meltwater Signatures and Mapping}

Meltwater can accumulate within the firn in the form of aquifers and subsurface lakes at depths of a few tens of meters, while refreezing can produce extensive ice layers at shallower depths \cite{hubbard2016massive, montgomery2020hydrologic, frank2026mysterious}. These features can produce strong dielectric contrasts with respect to the surrounding firn, detectable in Ground Penetrating Radar (GPR) and airborne sounder data. Signal attenuation may also occur below water-bearing regions due to electromagnetic absorption.\\
BIOMASS TomoSAR is side-looking and reconstructs the vertical backscatter distribution rather than direct reflections from interfaces, with much lower resolution than GPR or airborne sounders \cite{banda2026eusar}. Consequently, subsurface water and ice related features are expected to appear as scattering transitions or changes in backscatter, potentially mixed with contributions from the air/firn interface due to limited vertical resolution.\\
Polarization can help distinguish scattering contributions: volume is generally more prominent in VH/HV, whereas relatively smooth interfaces may dominate HH/VV and enhance the detectability of meltwater-related features. When such interfaces dominate the backscatter, TomoSAR phase center may shift from the surface towards the depth of the subsurface feature, providing an estimate of its location without necessarily capturing its full vertical extent.

\section{Area and Datasets}
\label{sec:dataset}

\subsection{Riiser-Larsen Ice Shelf}

The Riiser-Larsen Ice Shelf (RLIS), located in Queen Maud Land, East Antarctica (72°40'S, 15°00'W), is a floating ice mass harbored by East Antarctic outlet glaciers \cite{dirscherl2021seasonal, arthur2022large}. Its surface topography is generally smooth, with undulations and local depressions related to ice flow and surface morphology. Recurrent supraglacial lake activity has been documented, with lake formation preferentially occurring in surface depressions. Lake distribution and extent vary in response to environmental conditions and local topography. According to the Sentinel-based PolarLakes maps \cite{baumhoer2025polarlakes, baumhoer2026polarlakes}, maximum lake extent decreased markedly after the 2019/20 melt season, from several tens to only a few square kilometers. These observations are consistent with reports by \cite{dirscherl2021seasonal} for the same season.\\
GPR acquisitions from 2025 revealed clear signals resembling liquid water. Borehole investigations at the corresponding location confirmed the presence of a water-filled cavity at approximately 30 m depth \cite{frank2026mysterious}. With these and additional observations, researchers suggest that buried water may persist over multiple seasons, potentially undergoing episodic drainage \cite{frank2026mysterious}.

\subsection{Datasets}

\begin{table}[ht]
\centering
\caption{Parameters of the processed BIOMASS stacks.}
\renewcommand{\arraystretch}{1.2}
\begin{tabular}{|c|c|c|c|c|}
\hline
\textbf{Phase \& Dates} & \textbf{Swath} / $\theta$ & $\mathbf\rho_z$ [m] & \textbf{HoA} [m] & $N_a$ \\ \hline
 C2-1, Jul 3/18, 2025 & S1 / 23° & 23 & 100 & 6 \\ \hline
 C2-2, Jul 21/Aug 5, 2025 & S1 / 23° & 23 & 100 & 6 \\ \hline
 C4, Oct 7/22, 2025 & S2 / 25° & 63 & 325 & 6 \\ \hline
 TO, Jun 4/22, 2026 & S3 / 28° & 63 & 325 & 7 \\ \hline
\end{tabular}
\label{tab:biomassparams}
\end{table}

Tab.\ref{tab:biomassparams} reports main parameters of the BIOMASS stacks processed from commissioning (Cx) and tomographic (TO) phases \cite{scipal2026biomass}, including nominal incidence angle $\theta$, vertical resolution $\rho_z$, height of ambiguity (HoA) and number of acquisitions $N_a$. Common features are the full polarimetry, 3-days repeat, fairly regular baselines, single look azimuth-slant range resolution of about \mbox{$8 \times 25$ m$^2$}. Stacks are pre-processed with BIOMASS Processing Suite (BPS) \cite{BPS}, starting from raw data and applying L1a processing, i.e., focusing, polarimetric calibration and single-acquisition ionosphere correction. L1c stack processing includes coregistration and multi-squint InSAR ionosphere calibration. The latter further improves over plain coregistration, although ionosphere is not as harming as in other scenarios \cite{banda2026eusar}. Assuming scene reciprocity, VH/HV are averaged to gain 3 dB over noise.\\
Ad-hoc processing described in Section \ref{sec:methodology} is applied to estimate SKPD and PL, SKPD surface phases are applied before computing coherence over a \mbox{300 m} ground window, which gives $>200$ independent looks. Phase center depth is mapped from Fourier TomoSAR normalized for each azimuth-range cell, as described in Section \ref{sec:methodology} and further in Section \ref{sec:results}. Depth map is 3D geocoded and corrected by scaling depths with $n=1.7$ refraction index \cite{sharma2012estimation, banda2015single}.\\
PolarLakes \cite{baumhoer2025polarlakes} is used for reference, particularly the total recurrence map, indicating the cumulative number of bi-weekly detections of supraglacial lakes during the melt season (November–March) over the 2016/2024 period. The map is provided as a raster with 10 m posting, in Antarctic Polar Stereographic projection.

\section{Experimental Results}
\label{sec:results}

\subsection{Surface Steering and TomoSAR}

\begin{figure}[ht!]

		\centering
		\includegraphics[width=1\linewidth]{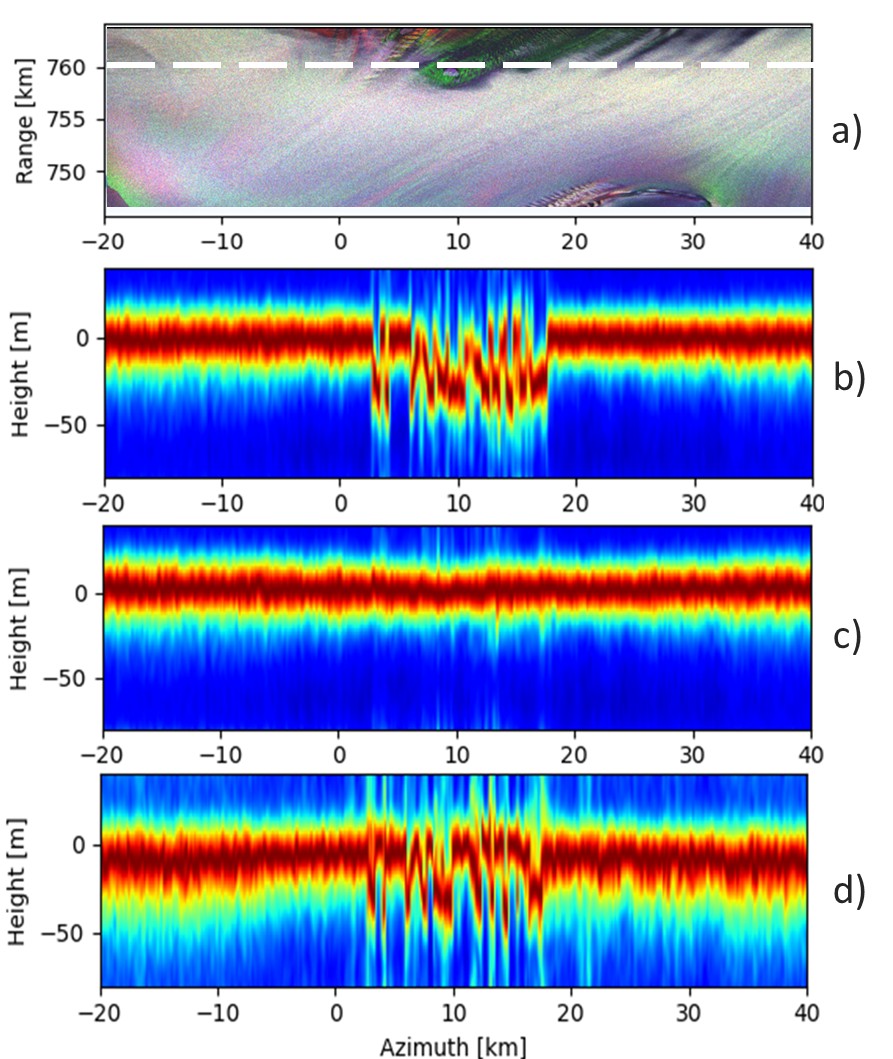}
		
		\caption{a) RLIS Pauli: R=HH-VV, G=HV+VH, B=HH+VV. b-d) C2-1 Fourier TomoSAR for the dashed-line transect of (a): HH calibrated with SKPD (b); HH calibrated with PL (c); VH calibrated with SKPD (d). Refraction not yet corrected.}
		\label{fig:steer}
		
\end{figure}

Fig.\ref{fig:steer} reports Pauli composite for the area with former supraglacial lakes, corresponding to the triangularly shaped area on top of the plot. C2-1 Fourier TomoSAR slices are also reported for HH and VH, whereas VV shows a behavior similar to HH. By comparing HH and VH calibrated with SKPD, two distinct scattering areas can be identified: one characterized by surface and homogeneous subsurface volume; another corresponding to the former lakes, with HH signal more height-localized and distributed with relatively more lateral continuity towards the bottom of subsurface scattering, while VH spans from zero height level to subsurface, with more signal close to surface. This is consistent with expected meltwater-related structures described in Section \ref{sec:methodology} and resembles GPR observations from \cite{frank2026mysterious}, where the lake boundary is sensibly brighter than the air/firn interface.\\
Regarding surface steering, it can be appreciated that SKPD-based calibration consistently reconstructs surface level while preserving subsurface scattering features across different polarizations. Instead, PL-based calibration flattens out scattering transitions between surface and subsurface, potentially making the extra surface steering more difficult. The surface/subsurface separation achieved through SKPD, together with the two-step procedure described in Section \ref{sec:methodology}, yields consistent calibration and steering, without issues of identifiability between the surface and subsurface mechanisms.

\begin{figure}[ht!]

		\centering
		\includegraphics[width=1\linewidth]{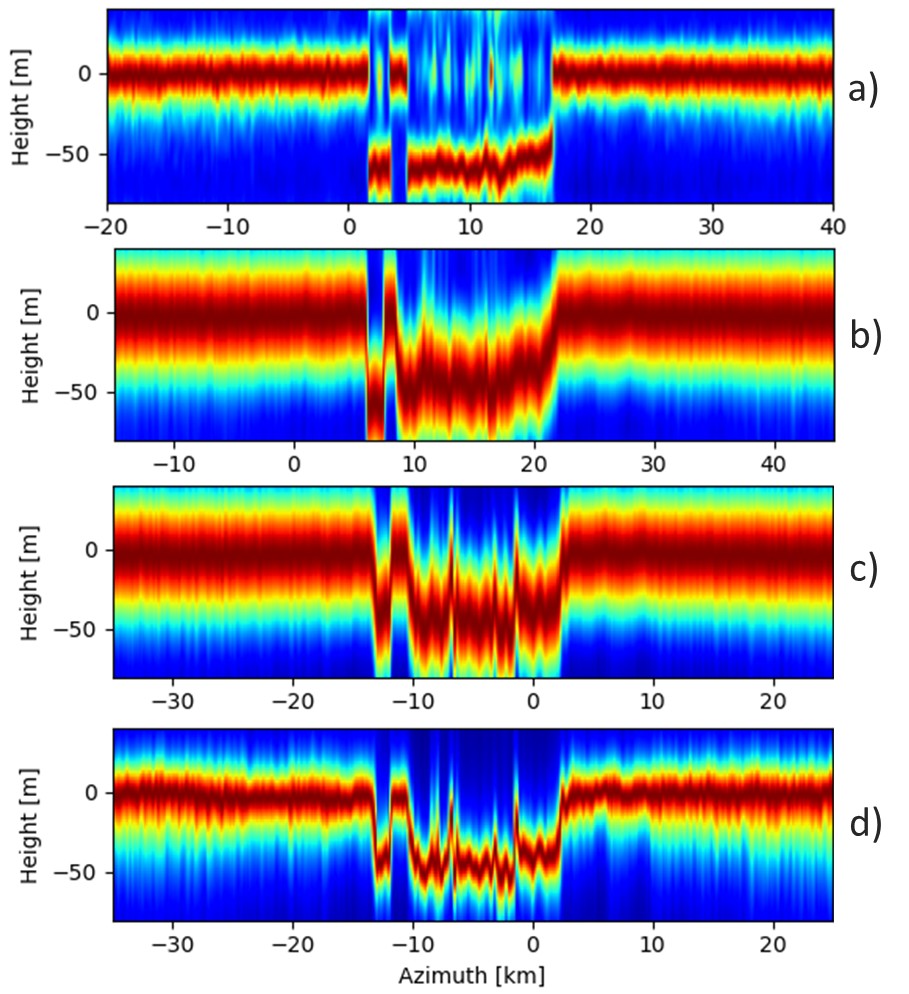}
		
		\caption{TomoSAR for approximately the same location of Fig.\ref{fig:steer}. Azimuth axis is representative of the different local swaths. Refraction not yet corrected. a) C2-2 Fourier HH. b) C4 Fourier HH. c) TO Fourier HH. d) TO Capon HH.}
		\label{fig:tomoCAPON}
		
\end{figure}

\noindent Fig.\ref{fig:tomoCAPON} reports HH TomoSAR slices for the other stacks at approximately the same location of Fig.\ref{fig:steer}. Similar scattering areas and subsurface features can be observed. Temporal redistribution of subsurface scattering in correspondence of former supraglacial lakes is evident, with phase center showing improved lateral continuity at lower depth in C2-2, lifting up a bit in C4 and remaining substantially stable in TO. VH tomograms are not reported but remain consistent with the observations of Fig.\ref{fig:steer}. Whereas ineffective for C2 due to the spectral shift effect \cite{gatelli1994wavenumber}, Capon can partly recover the much coarser resolution of TO (likewise for C4), though not improving the phase center localization. Based on this, the subsurface signature location is identified as corresponding to the position of the main lobe in HH Fourier TomoSAR. Mapping is computed only for areas where the difference of $z_C$ between PL and SKPD phases exceeds average PL phase center location (see Fig.\ref{fig:steer}).

\subsection{Meltwater Signatures Mapping}

To investigate the relationship between TomoSAR and PolarLakes, phase center depth maps are generated as described in Sections \ref{sec:methodology}-\ref{sec:results}A, including correction of propagation in ice, and projected to PolarLakes local frame, focusing on former supraglacial lakes area. Averaging to 300 m ground window is applied also to PolarLakes total recurrence map, to achieve a fairer comparison.

\begin{figure*}[ht!]

		\centering
		\includegraphics[width=1\linewidth]{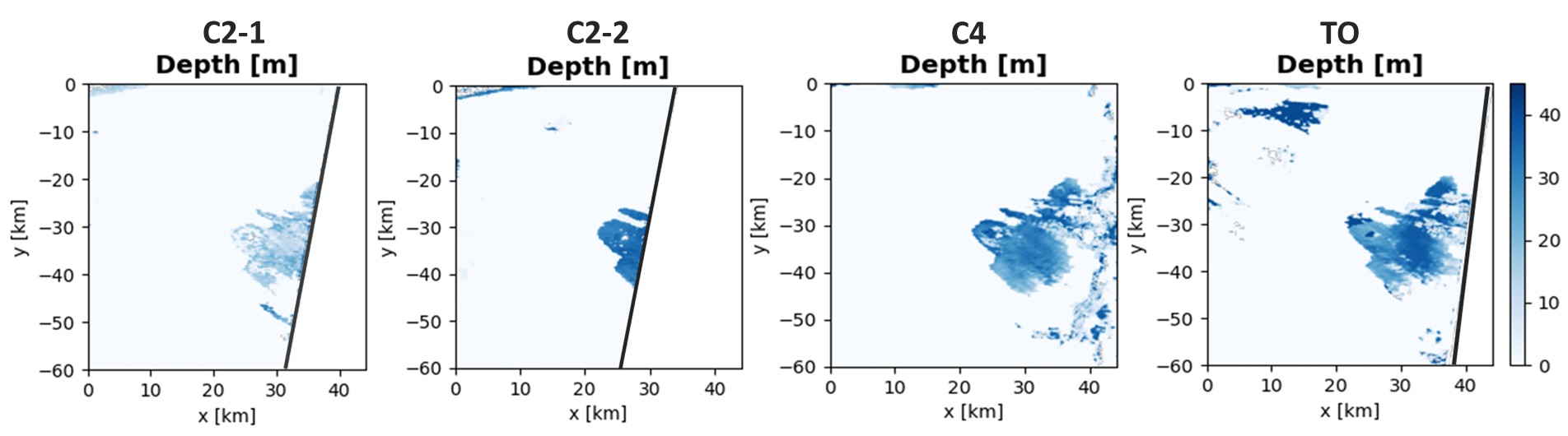}
		
		\caption{From left to right: phase center depth as detected in C2-1, C2-2, C4, TO BIOMASS TomoSAR stacks. Black lines delimit swath extent.}
		\label{fig:lakemaps}
		
\end{figure*}

\begin{figure}[ht!]

		\centering
		\includegraphics[width=.6\linewidth]{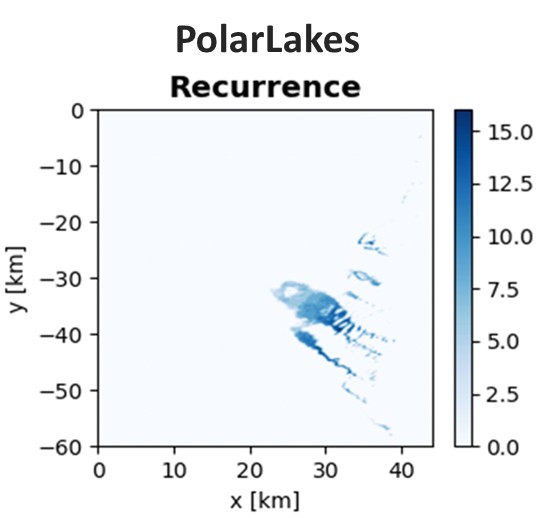}
		
		\caption{Surface lakes total recurrence map from PolarLakes.}
		\label{fig:lakemaps2}
		
\end{figure}

\begin{figure}[ht!]

		\centering
		\includegraphics[width=1\linewidth]{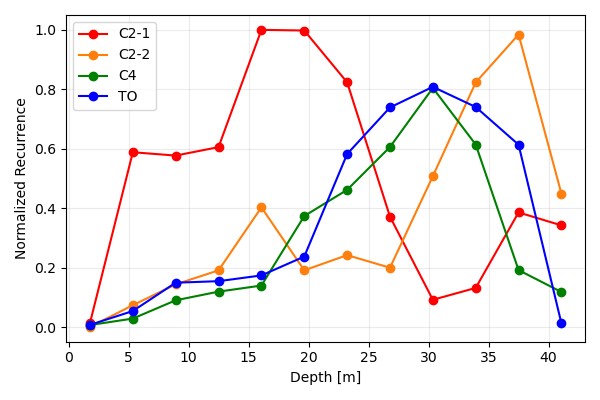}
		
		\caption{Normalized mean recurrence as a function of depth for all four BIOMASS-derived maps.}
		\label{fig:recmean}
		
\end{figure}

\begin{figure}[ht!]

		\centering
		\includegraphics[width=1\linewidth]{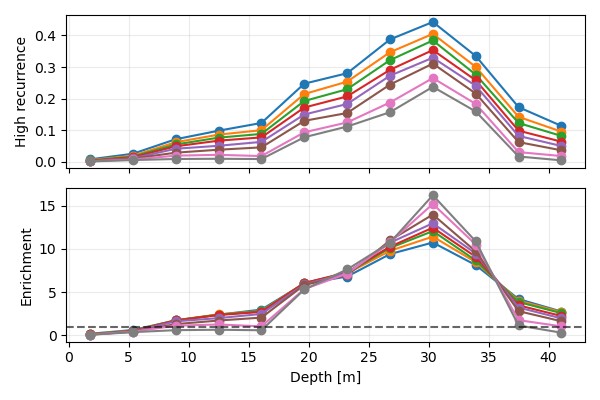}
		
		\caption{Fraction of high recurrence points (top) and enrichment (bottom) as a function of depth for C4. Colors from blue to gray represent increasing threshold on recurrence from 6\% to 50\% of the maximum.}
		\label{fig:thresh}
		
\end{figure}

\noindent Fig.\ref{fig:lakemaps}-\ref{fig:recmean} report the spatial distribution and evolution of the subsurface signature, as well as its association with historical lake recurrence, through normalized mean recurrence as a function of depth. The analysis is restricted to points with depths greater than 5 m, excluding the very shallowest region. The signature matches the overall morphology but extends beyond previously mapped surface lake boundaries, suggesting that subsurface activity is not necessarily confined to these locations. BIOMASS-derived signature reaches maximum mean recurrence approximately between 18 and 37 m depth, with a progressively narrower distribution over time. Although differences may be partly attributable to variations in TomoSAR coverage and configuration, it is worth highlighting the different seasonal conditions: C2 and TO were acquired during austral winter one year apart, whereas C4 took place in the austral spring after C2, immediately before the onset of the melting season.\\
To further investigate the depth-recurrence relation, a recurrence threshold from 6\% to 50\% of the maximum is varied to classify points into high and low recurrence events. For each threshold, the fraction of high recurrence points is computed for each depth interval; in addition, the enrichment factor is computed as:

\begin{equation}
	e_d(t)=f_d(t)/f_g(t)
\end{equation}

\noindent where $f_d(t)$ is the fraction of high recurrence points for depth $d$ and threshold $t$, and $f_g(t)$ is the fraction of high recurrence points over total observations, regardless of point depth. Richness values above unity indicate a greater proportion of high recurring locations within the corresponding depth interval. Results are reported in Fig.\ref{fig:thresh} for C4 and with similar results for other stacks further support the robustness of the observed depth-recurrence trend.

\section{Discussion}
\label{sec:discussion}

The spatial match between historical lake recurrence and persistent BIOMASS subsurface signatures, along with recent marked reduction in surface lake extent, could be consistent with long-term modifications of firn structure, possibly caused by repeated meltwater infiltrations. The extension beyond historical lake boundaries suggests that subsurface activity is not necessarily confined to these exact locations.\\
The depth and lateral continuity of the signatures are compatible with Radar signals associated to hydrologically modified firn. An example of Antarctic firn aquifer is provided by \cite{montgomery2020hydrologic}, and \cite{frank2026mysterious} suggests that latent heat released during refreezing can progressively modify firn thermal state, favoring the formation of water-filled cavities; the one mapped at RLIS in 2025, the same year of the first three BIOMASS stacks processed in this study, was confirmed by borehole investigations \cite{frank2026mysterious}. A mix of refrozen and residual liquid water also represents a possible interpretation.\\
Nevertheless, downward-looking Radar and TomoSAR rely on different measurement principles, and TomoSAR signatures cannot be uniquely interpreted at this stage. The evolution from a more discontinuous to a less diffuse scattering over time may reflect changes in relative contributions of different subsurface scattering components, but this should not necessarily be interpreted as physical migration of a waterfront without further validation.

\section{Conclusions}
\label{sec:conclusions}

In this work, P-band TomoSAR datasets processed from different BIOMASS mission phases reveal laterally coherent subsurface features, in correspondence of historical surface lake activity at the Riiser-Larsen Ice Shelf. Assuming refraction index in ice of $n=1.7$, the effective scattering occurs at depths ranging from about 18 to 37 m and can be mapped across four stacks, between July 2025 and June 2026.\\
The spatial association with historical supraglacial lake recurrence and temporal persistence may indicate long-term hydrological modification of the underlying firn, with potential relevance to its response to future surface melting and, more broadly, to ice shelf stability. Although depth and continuity of the signatures are compatible with documented meltwater-related Radar structures and water-filled cavities have been recently mapped at RLIS \cite{frank2026mysterious}, observations cannot be uniquely linked yet to the presence of liquid water or an aquifer.\\
This research highlights the significant potential of BIOMASS for investigating unresolved aspects of subsurface hydrology in Antarctic ice shelves, while also providing broader insights about internal ice structure. By addressing an existing gap in satellite remote sensing, BIOMASS could support a wide range of glaciological applications in cryosphere science.\\
Future work will involve further assessment and validation at RLIS and other sites, possibly in collaboration with researchers from the glaciological community, alongside physical modeling of scattering mechanisms \cite{banda2015single, fischer2019modeling}.

\raggedright
\bibliographystyle{IEEEtran}
\bibliography{bibliography}

\vfill

\end{document}